# Results of numerical integrations of the orbits of fictive asteroids in vicinity of 2:1, 3:1 and 3:2 resonances.

**Eva Plavalova**[1] **, Alexey Rosaev**[2]

[1]Mathemailcal institute, Slovak academy of Science, Bratislava, Slovakia
email: plavalova@mat.savba.sk, plavalova@komplet.sk
[2]*Regional Scientific and Educational Mathematical Center Centre of Integrable System, Yaroslavl, Russia,* email: hegem@mail.ru

## Introduction

In this paper we present our results of numerical integrations of orbits of fictive massless particle in vicinity of resonance. Our goal is to study the dependences of period (frequency) of resonance perturbations and the width of the resonance on mass and orbital eccentricity of perturbing planet and on the initial difference of longitude between test particle and perturbing planet. The main attention is paid to planar restricted three body problem, but some computations are done for a spatial case.

There are very many papers studying the resonant motion, but most of them deal with the libration solution (see [1], [2] and references herein). In this paper we study the case of resonance argument circulation because even the weak resonant perturbation can have a notable effect on the small bodies motion.

The main target of this preprint is to give data for future theoretic study.

## Methods and problem setting

In this paper we present our results of numerical integrations of orbits of several fictive massless particles in vicinity of the resonance with the eccentricity 0.1 and with the step in initial semimajor axis equal 0.005 AU. However, the average values of semimajor axes are notably differ from initial (due to the planet perturbation). Additionally, we made integration with inclination of minor body i=5 degrees, but inclination has negligible effect on the main amplitudes and frequencies, so it is not discussed here. Also, we varying the mass and eccentricity of the perturbing planet: m=0.1mj, m=0.01mj, m=0.5mj and ep=0, ep=0.048. All orbits starts from perihelion, when we account the planet's orbit eccentricity, it coincide with planet perihelion. We use standard package Mercury [3] and the Radau method [4] in our integration. In this preprint we report only about clones on the orbits close but not in the resonance i.e. orbits with circulated resonant argument.

The evolution of semimajor axis is approximated by our method [5]. To obtain the approximation we use a minimum of the function:

$$\sigma_k = \sqrt{\frac{1}{n-2}\sum_{i=1}^{n}(X_i - X_{iapprox})^2} \qquad (1)$$

Here n is number of samples, $X_i$ is the value obtained by the numerical solution, $X_{iapprox}$ is the approximating value. An example of our approximation is given in Fig.1.

For the determination of the width of resonance we use the well known fact that perihelion in resonance has reverse rotation. Despite the method is rough, it allowed us to describe the results of numerical integration. It is sufficient to verify and illustrate our analytical model.

To test our theoretic model, we made integration of the three body problem (Sun-Jupiter-Asteroid) with complete perturbation of planet. The massless test particle (asteroid) orbited in different distance d close to resonance with Jupiter (i.e. with semimajor axis $a_{res}$+d).

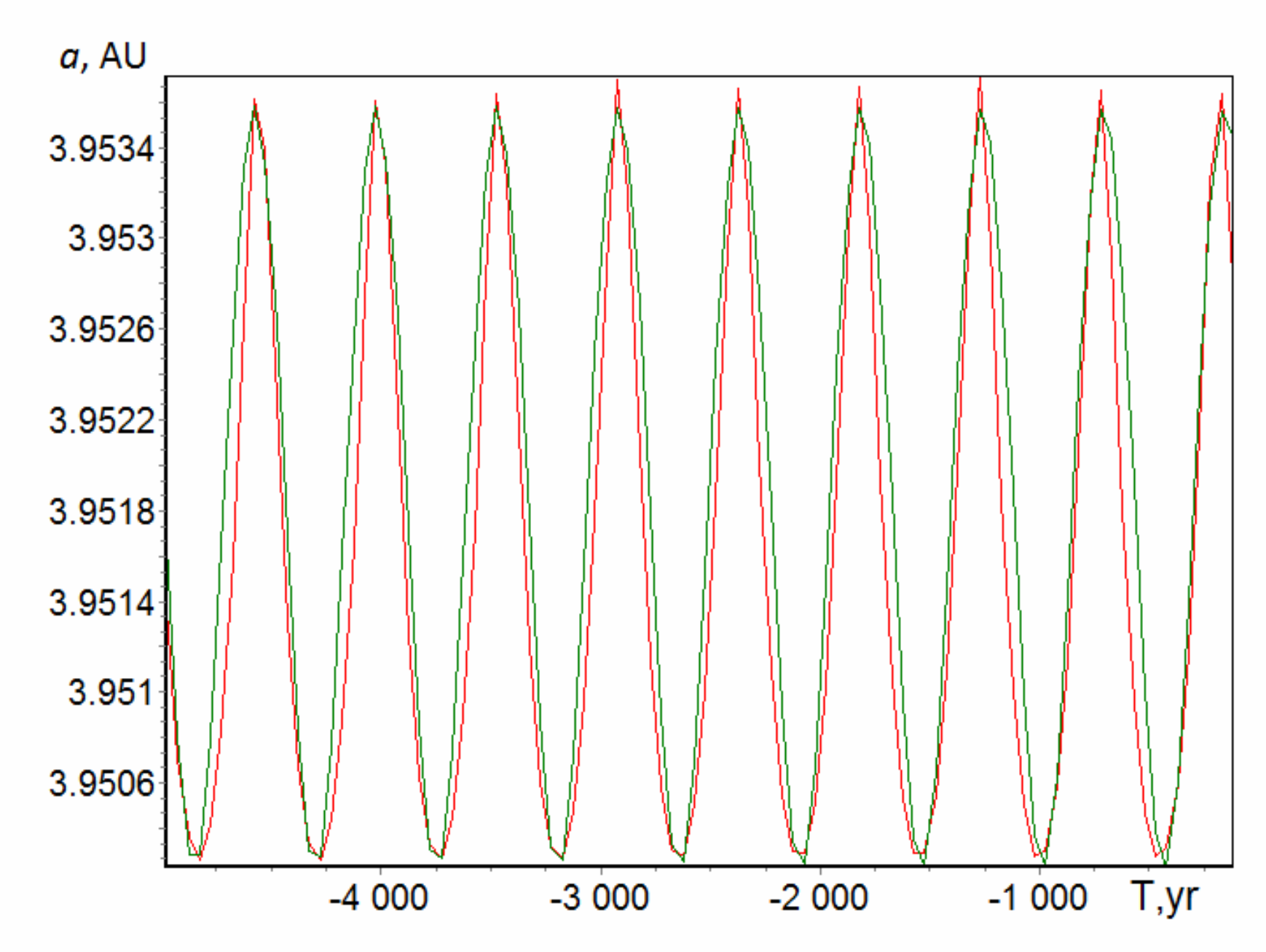

Fig. 1. The semimajor axis evolution (red) and its approximation (green). Planet mass m=0.01mJ, planet eccentricity is ep=0, test particle orbital elements are e=0.1, i=0.

In order to make a comparison with the results of numerical integration, it is necessary to move from changes in the radius vector to the evolution of the elements of the orbits (in the planar case a and e):

$$\delta r(t) = (1-e^2)\delta a - 2ea\delta e$$

The dependence between $\delta a$ and $\delta e$ is given by the equation (8.35) from the book by Murray and Dermott [1].

This means that the frequency (period) of the resonant disturbance is the same for the semmajor axis and eccentricity. Therefore, we can use the orbital element that is most convenient to describe the resonance. It follows from the above equations that the increment of eccentricity $\delta e$ makes the main contribution to $x = \delta r$ only at small e. However, if we consider the relative values $\delta a / a$, $\delta e / e$, then obviously the changes in eccentricity are the greatest. With the circular orbit of the disturbing planet (Jupiter), changes in the semimajor axis allow us to give a better performance (Fig. 2.).

The variations of the semimajor axis and eccentricity have the same period and there are not any long periodic (secular) perturbations.

The behavior of test particle in the planar circular restricted three body problem is described by four variables. However the position on the orbit can be removed by averaging and the value of $\dot{\varpi}$ is constant. Therefore all dynamical behavior can be controlled by two parameters: eccentricity and semimajor axis.

### 2:1 resonance

The value of semimajor axis corresponded with exact 2:1 commensurability with Jupiter is $a_{res}$=3.27769 AU. As it is evident in the fig 3, the frequency of the resonant perturbations decreases in approach to resonance when the amplitude increases.

In result we have obtained the dependence of frequency of perturbation on distance to resonance in a fine agreement with our model (Fig.4). **The interval between 3.265 and 3.345 AU corresponds with libration solution and is not considered in present paper.** After that we have repeated our integrations and approximation with the actual eccentricity of Jupiter (e=0.048) and obtain the same values of frequency and amplitude and obtain the same dependence.

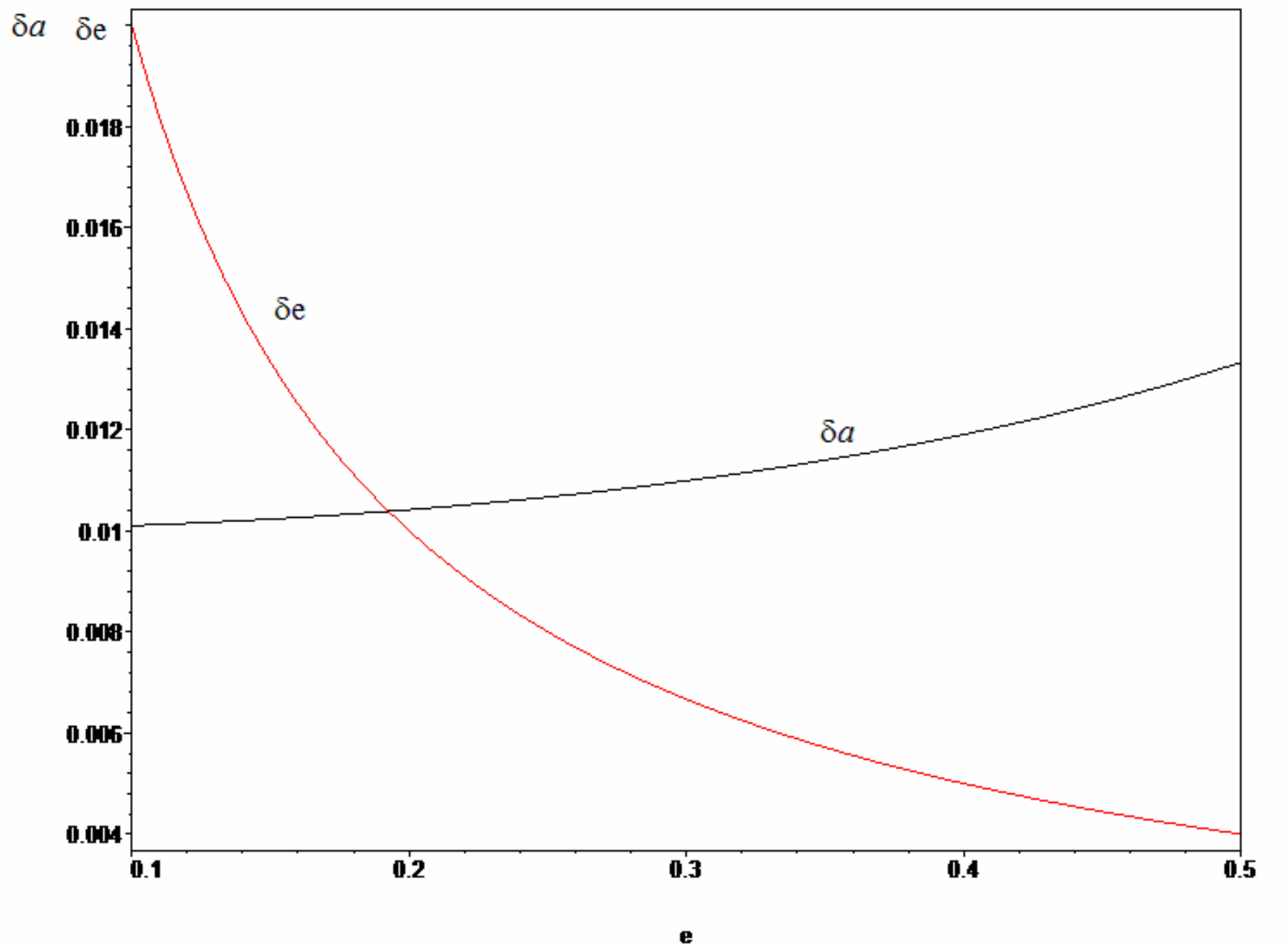

Fig. 2. The contribution $\delta a$, $\delta e$ in $\overset{e}{\delta r}$ as a function of eccentricity

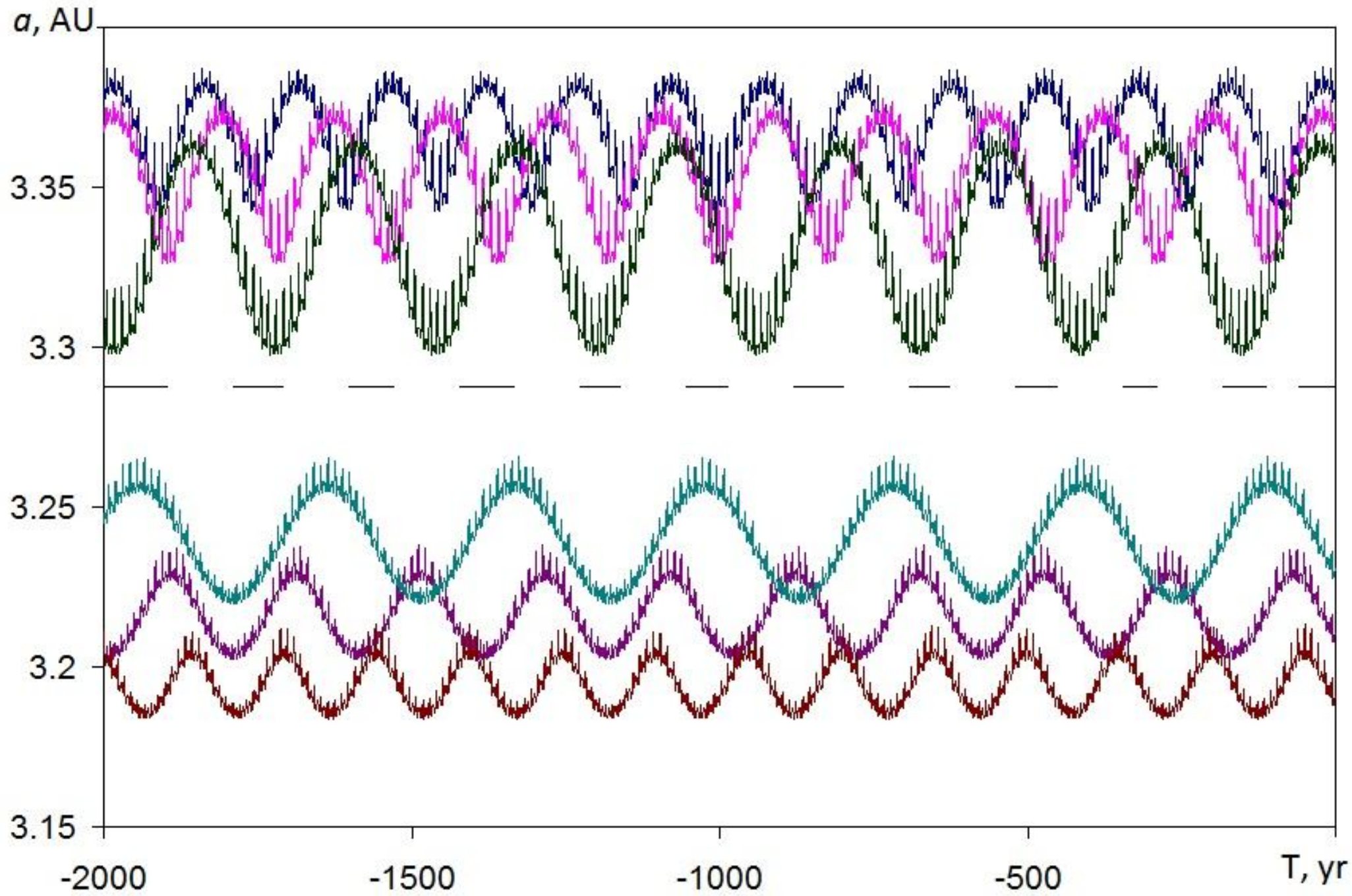

Fig.3. The perturbations corresponded 2:1J resonance for different values of semimajor axis. The nominal position of resonance is marked by dashed line.

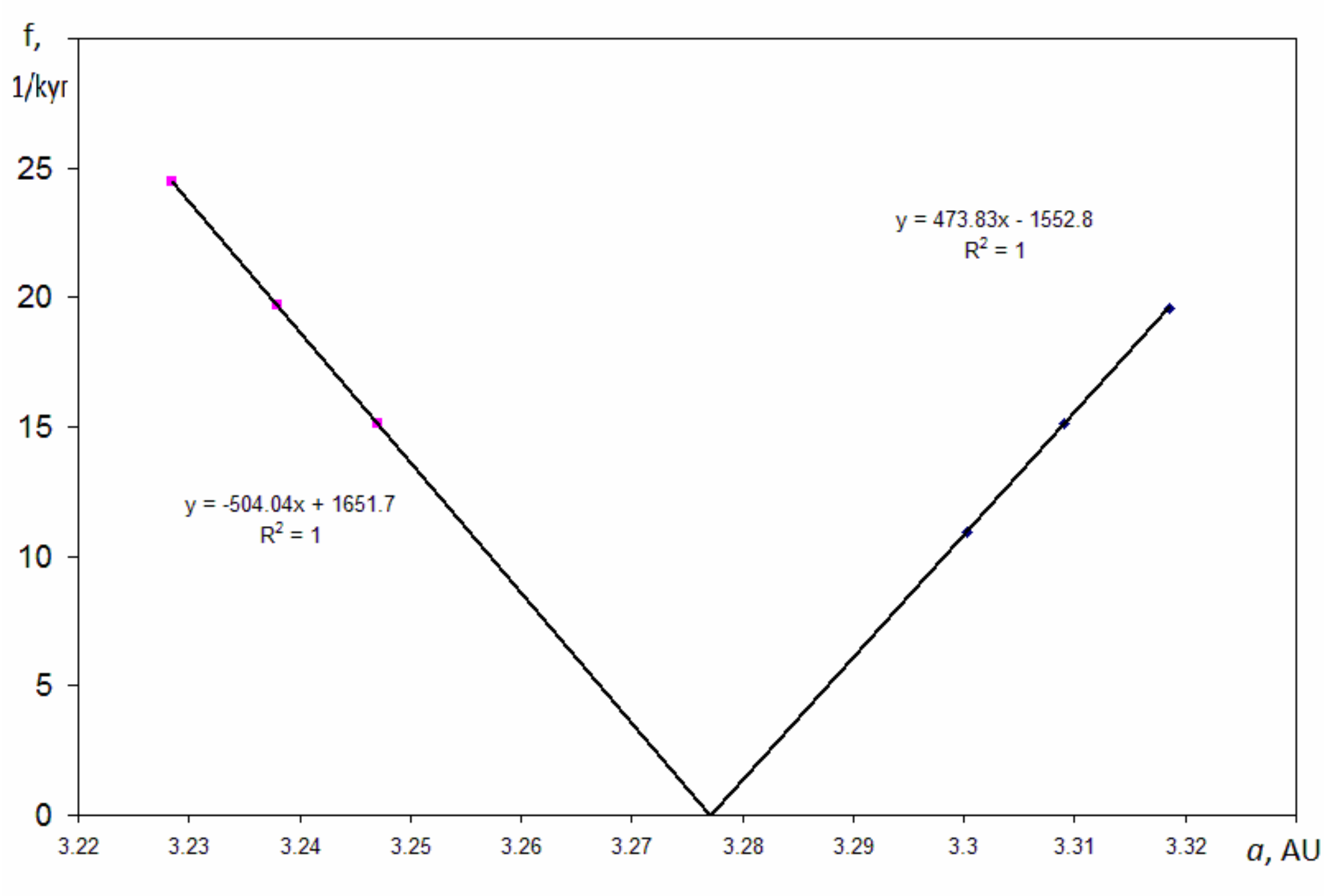

Fig.4. The dependence of the resonant perturbation frequency on the semimajor axis near to 2:1J resonance. The nominal position of resonance is equal 3.278 AU. M ass of perturbing planet m=0.1 mJ.

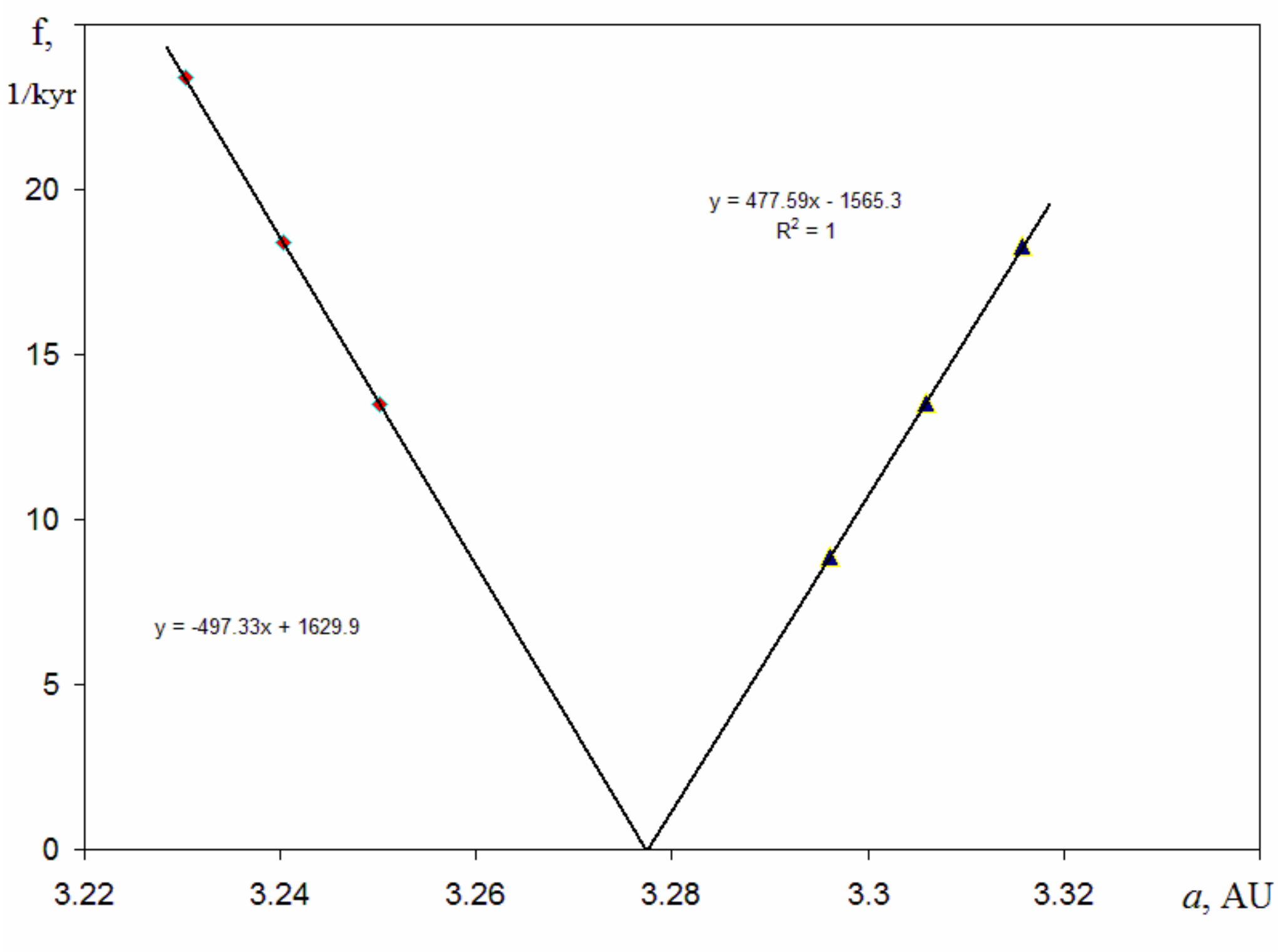

Fig.5. The dependence of the resonant perturbation frequency on the semimajor axis near to 2:1J resonance. The nominal position of resonance is equal 3.278 AU. M ass of perturbing planet m=0.01 mJ.

Table 1. Result of approximation of numerical integration of fictive orbits
in vicinity 2:1 resonance for eJ=0, planet mass is equal m=0.1J

| File | approximation | sigma |
|---|---|---|
| 217 | a=3.228+0.001854cos(0.02448t+0.380) | 0.00021 |
| 218 | a=3.238+0.002496cos(0.01968t+ 0.101) | 0.00029 |
| 219 | a=3.247+ 0.003333cos(0.0151t+6.225) | 0.00029 |
| | | |
| 221 | a=3.300+ 0.00449cos(0.01093t+ 2.988) | 0.00035 |
| 222 | a=3.309+ 0.0033cos(0.01513t+ 2.997) | 0.00037 |
| 223 | a=3.3185+ 0.0025cos(0.0196t+3.0) | 0.00030 |

Additionally, we integrate and approximate the evolution of some real asteroids near 2:1 resonance (Table 4, Fig. 6). We integrate ten largest asteroids inside 2:1 resonance and all known asteroids outside it. The dispersion in Fig.6 can be explained by different eccentricities and inclinations of real asteroids relative the modeled ones.

Table 2. Result of approximation of numerical integration of fictive orbits
in vicinity 2:1 resonance for ej=0.0, planet mass is equal m=0.01J

| File | approximation | sigma |
|---|---|---|
| 217 | a=3.2303+0.000196cos(0.02338t+ 0.00) | 0.000026 |
| 218 | a=3.2402+0.00025cos(0.01840t+ 0.10) | 0.000030 |
| 219 | a=3.2501+0.00035cos(0.01351t+ 0.10) | 0.000050 |
| | | |
| 221 | a=3.2961+0.000572cos(0.008882t+2.991) | 0.000067 |
| 222 | a=3.3059+0.00035cos(0.01355t+ 3.000) | 0.000059 |
| 223 | a=3.31583+0.00026cos(0.0183t+3.50) | 0.000039 |

For case perturbing mass m=0.1mJ the equations are:

F=473.83x - 1552.8

F=-504.04x + 1651.7

The corresponding intersection point ordinate is $a_{res}$=3.2770205 AU

For case perturbing mass m=0.01mJ the equations are:

F=-497.33x + 1629.9

F=477.59x - 1565.3

The corresponding intersection point ordinate is $a_{res}$ =3.2773971 AU

Table 3. Results of approximation of numerical integration of fictive orbits in vicinity 2:1 resonance (ej=0.048).

| file | *a*, AU | Eccentricity approximation | sigma |
|---|---|---|---|
| 214 | 3.210 | 0.0850+ 0.0270cos(0.001080t+1.50) | |
| 215 | 3.220 | 0.0850+0.0260cos(0.001287t+1.33) | |
| 217 | 3.240 | 0.070+0.0130cos(0.00238t+1.56) | |
| | | | |
| 228 | 3.260 | 0.1940+0.0550cos(0.00277t+2.10) | |
| 229 | 3.350 | 0.1280+0.0401cos(0.00185t+2.01) | |
| 230 | 3.365 | 0.1250+0.0370cos(0.00149t+2.13) | |

Table 4. Result of approximation of numerical integration of the orbits of selected asteroids in vicinity 2:1 resonance $a_{res}$=3.27769

| #asteroid | approximation |
|---|---|
| 1295 | 3.3820+ 0.01746*cos(0.04990t+ 0.0) |
| 9398 | 3.3550+ 0.01824*cos(0.03686t+2.51) |
| 275408 | 3.2800+0.08686*cos(0.02075t+0.301) |
| | |
| 1229 | 3.2170+0.02304*cos(0.031932t+0.99) |
| 1956 | 3.2040+0.01575*cos(0.038878t+0.0) |
| 2342 | 3.2100+0.02000*cos(0.03620t+0.00) |
| 2722 | 3.2090+0.02184*cos(0.03650t+4.02) |
| 2848 | 3.2160+0.0240*cos(0.032532t+3.52) |
| 3766 | 3.2150+0.02112*cos(0.03350t+5.98) |
| 3785 | 3.2250+0.02688*cos(0.02929t+ 4.98) |
| 3797 | 3.2060+0.02208*cos(0.038323t+5.03) |
| 4814 | 3.2140+0.027295*cos(0.03453t+ 5.02) |
| 34801 | 3.240+0.02910*cos(0.02070t+ 5.00) |
| 117883 | 3.230+0.03162*cos(0.02532t+0.0) |

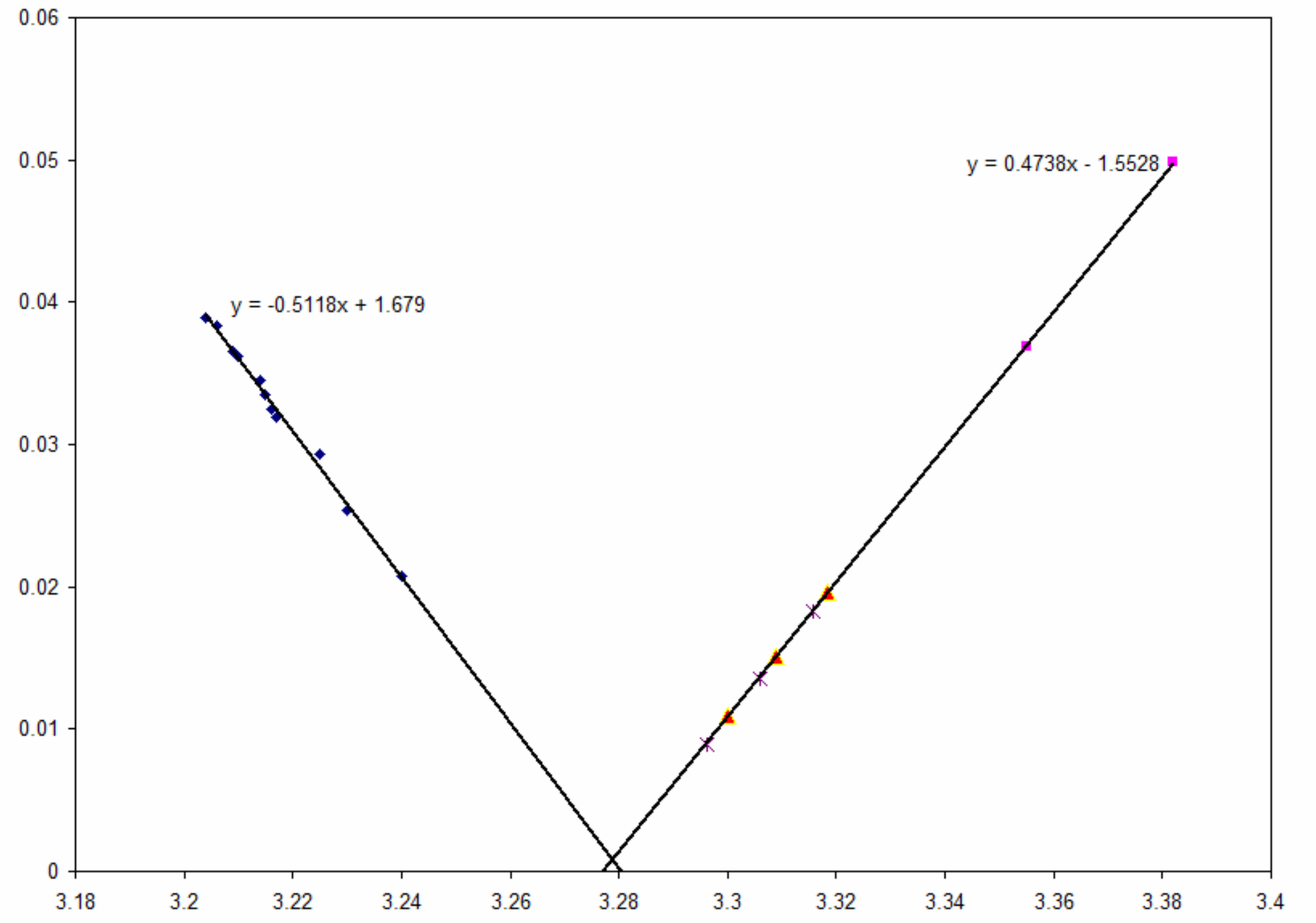

Fig.6. The dependence of the resonant perturbation frequency on the semimajor axis for selected real asteroids near to 2:1J resonance. The nominal position of resonance is equal 3.278 AU.

**The 3:2 resonance**

The value of semimajor axis corresponded with exact commensurability with Jupiter *ares* is slightly different by different authors. By the MD book [1] *ares* = 3.9703201, by paper [6] *ares* = 3.968972667. By our previous numeric integrations, the value, averaged over 800 kyr interval is *ares* = 3.97031924 very close to the MD value.

At the outer boundary, the resonance 3:2J with Jupiter can overlap with 4:3J resonance (4.2946415 AU). The rough estimation of the boundary is at 4.132481 AU. To avoid this, we use the reduced mass of Jupiter to detailed study 3:2 resonance. To exclude secular perturbations, we assume the eccentricity of Jupiter orbit is equal zero. We integrate 25 fictive asteroids with different values of initial semimajor axis.

As it is known, there are two kinds of motion close to resonance with the libration and the circulation of the resonant argument. Here we point the main attention to the study the circulation solution. In result we obtain the periodic perturbations in semimajor axis. The frequency of the perturbations increases with distance to exact resonance when the amplitude decreases (Fig.7). A similar behaviour takes place for asteroid eccentricity.

Then we quantitative determine the main amplitudes and frequencies of perturbations in semimajor axis. The obtained values are confirmed by Fourier analyses method. Finally, we approximate obtained values of frequency by straight lines (separate above and below the exact resonance).

We repeat our study with non-zero Jupiter eccentricity and different Jupiter mass. The results are the same (Tables 5-7, Fig.8-9). In the same time, the dependence of amplitudes of perturbations (A) on distance to resonance ( $\delta$ ) approximated by expression (Fig.10):

$$A = 0.00034/\delta$$

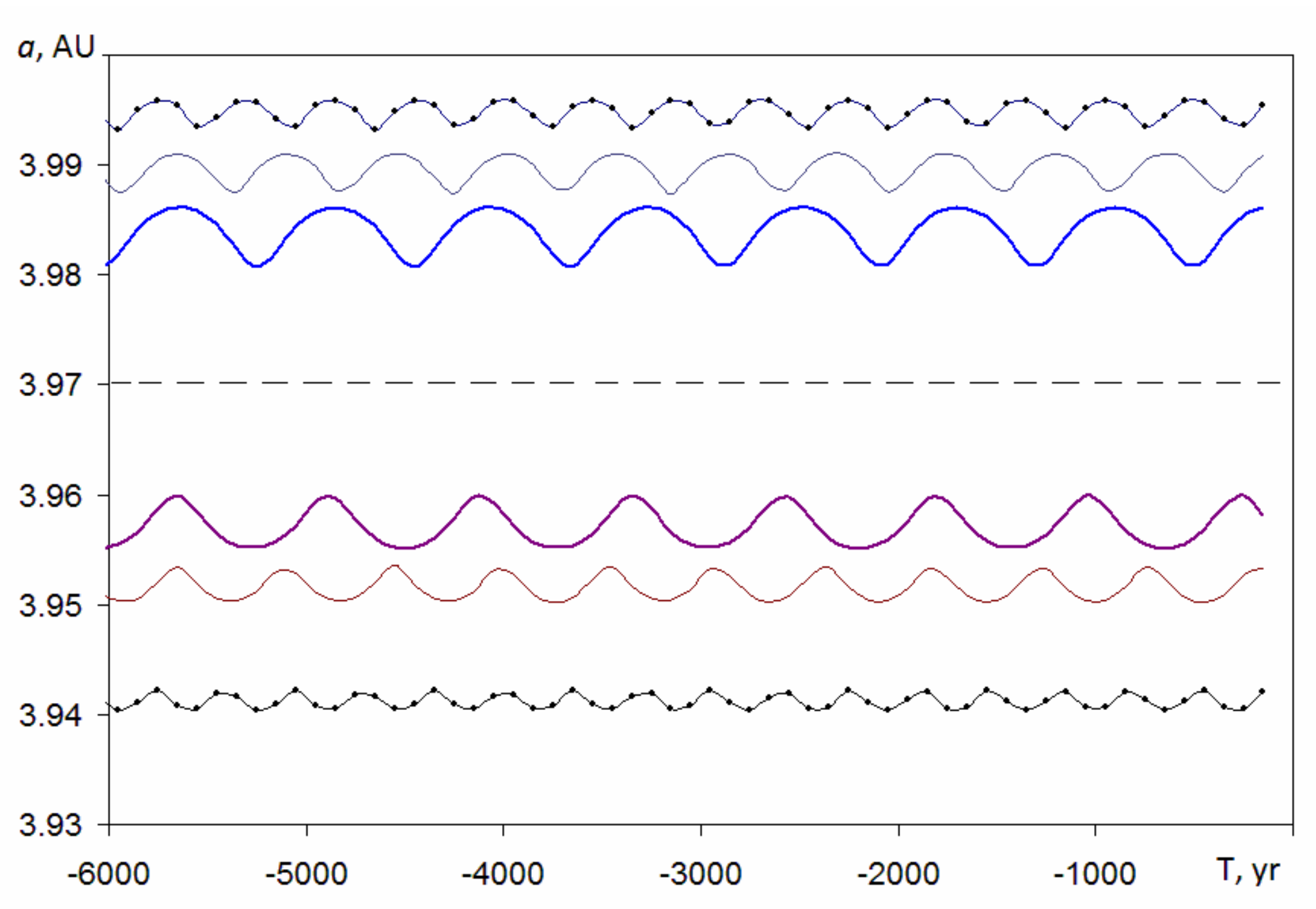

Fig.7 The dependence of the resonant perturbation frequency on the semimajor axis near to 3:2J resonance. The nominal position of resonance is equal 3.97 AU. All Jupiter perturbations are included.

Table 5. The results of approximation of evolution of semimajor axis of fictive orbits in vicinity of the 3:2 resonance. Planet eccentricity is equal eJ=0, planet mass is equal m=0.01J

| File | Semimajor axis approximation | sigma | Eccentricity approximation |
|---|---|---|---|
| 235 | a=3.9945+0.001365cos(14.456t+1.186) | 0.000308 | e=0.1009+0.00090cos(14.413t+4.221) |
| 234 | a=3.9892+0.001862cos(11.345t+1.055) | 0.000395 | e=0.1012+0.00116cos(11.368t+4.221) |
| 233 | a=3.9833+0.002828cos(08.062t+1.206) | 0.000345 | e=0.1016+0.00173cos(08.052t+4.323) |
| | | | |
| 227 | a=3.9575+0.002448cos(08.132t+2.165) | 0.000463 | e=0.0990+0.001545cos(08.099t+5.2) |
| 226 | a=3.9519+0.001682cos(11.388t+1.827) | 0.000334 | e=0.0994+0.001078cos(11.399t+5.025) |
| 225 | a=3.9413+0.000960cos(17.729t+1.709) | 0.000179 | e=0.0997+0.000666cos(17.747t+5.120) |

After that the dependence of frequency on semimajor axis in table 1 was approximated by least square method in standard software package. The approximating equations are:

$F=-593.16a + 2355.6$

$F=570.62a - 2264.9$

The corresponding intersection point ordinate is $a$=3.9702521 AU

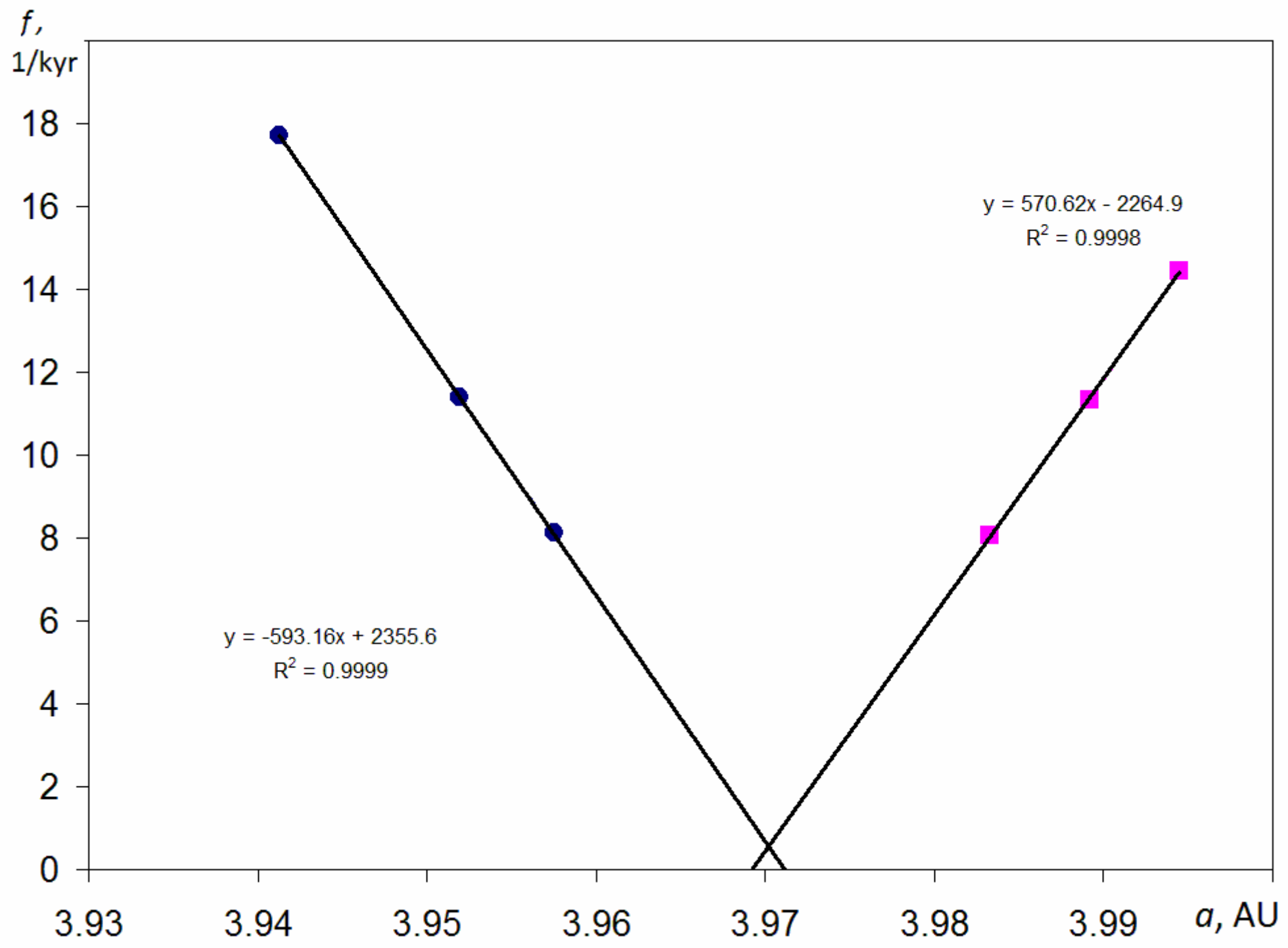

Fig.8. The dependence of the resonant perturbation frequency on the semimajor axis near to 3:2 resonance. The nominal position of resonance is equal 3.97 AU. Mass of perturbing planet is equal 0.01 of Jupiter mass.

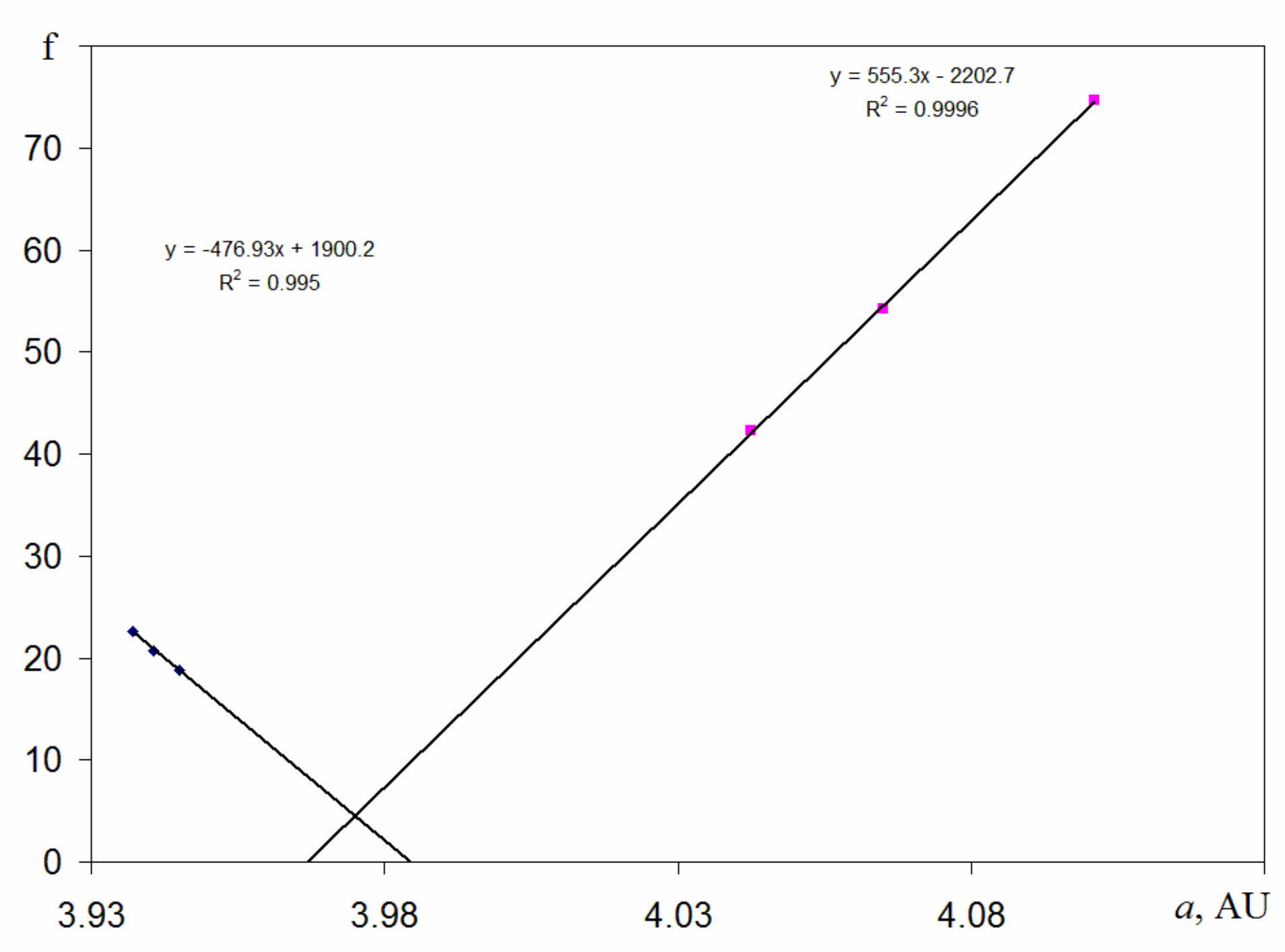

Fig.9. The dependence of the resonant perturbation frequency on the semimajor axis near to 3:2 resonance. The nominal position of resonance is equal 3.97 AU. Mass of perturbing planet is equal 0.5 of Jupiter mass.

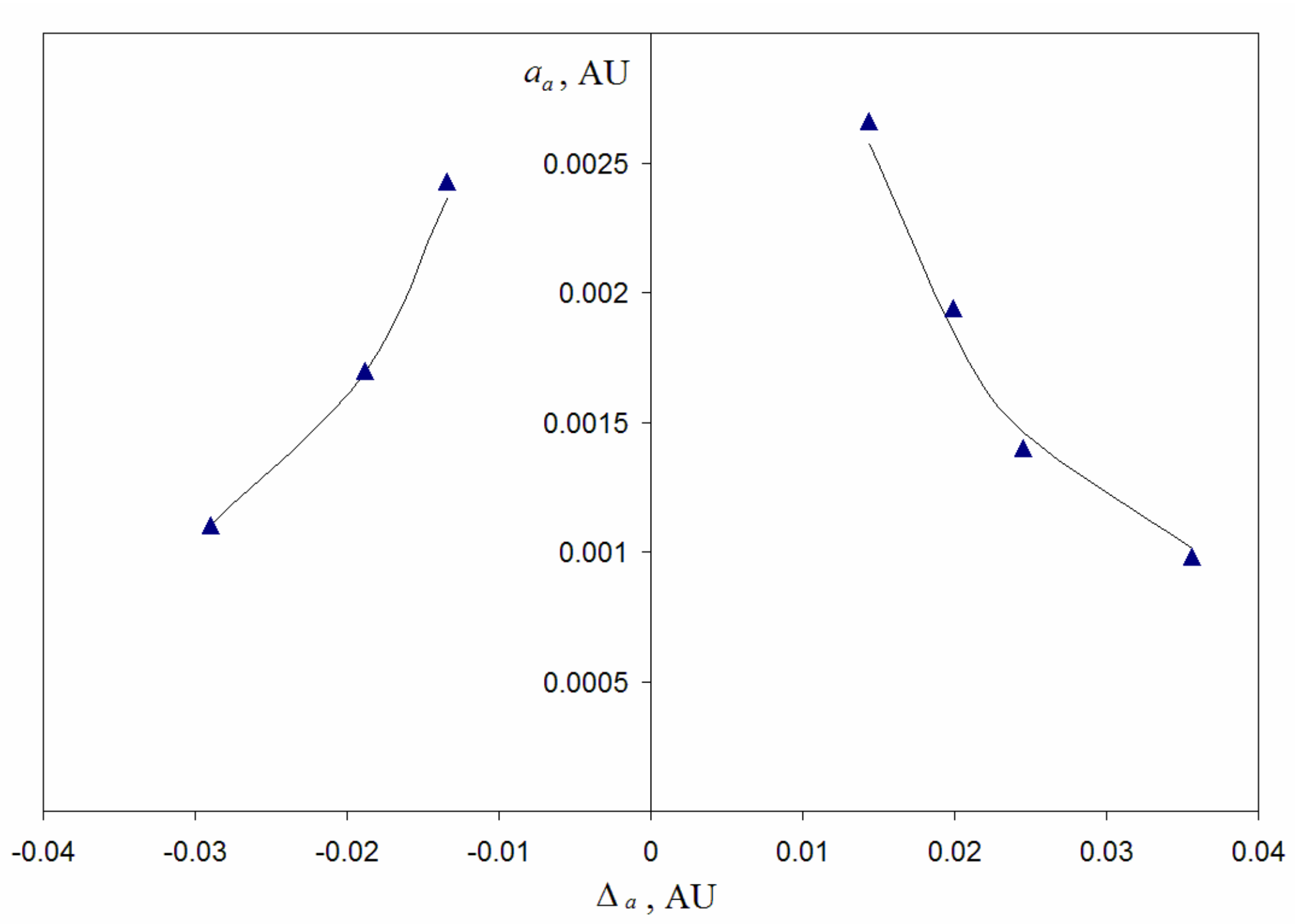

Fig.10. The dependence of amplitude of perturbations on distance to resonance

Table 6. The results of approximation of evolution of semimajor axis of fictive orbits in vicinity of the 3:2 resonance. Planet eccentricity is equal eJ=0, planet mass is equal m=0.1J

| File | Semimajor axis approximation | sigma | Eccentricity approximation |
|---|---|---|---|
| 241 | a=4.019+0.00672cos(29.06t+3.518) | 0.000938 | e=0.097+0.0042cos(28.942t+0) |
| 236 | a=4.012+0.00808cos(24.65t+3.770) | 0.001263 | e=0.098+0.0051cos(24.634t+0.503) |
| 240 | a=4.011+0.008225cos(24.16t+3.952) | 0.001196 | e=0.096+0.0058cos(24.002t+0.398) |
| 235 | a=4.0018+0.01089cos(19.15t+3.674) | 0.001370 | e=0.096+0.0063cos(19.246t+0.896) |
| 234 | a=3.9981+0.01275cos(16.327t+3.838) | 0.000988 | e=0.0965+0.0083cos(16.392t+0.896) |
| | | | |
| 226 | a=3.9388+0.01003cos(20.341t+5.976) | 0.001576 | e=0.108+0.00525cos(20.361t+2.988) |
| 225 | a=3.9299+0.00688cos(25.125t+6.150) | 0.000990 | e=0.107+0.0042cos(25.125t+2.988) |
| 224 | a=3.9257+0.00636cos(27.582t+6.101) | 0.000741 | e=0.107+0.00378cos(27.637t+2.988) |

The approximating equations are:
F=-550.38 $a$ + 2188.1
F=607.86 $a$ - 2414
The corresponding intersection point ordinate is $a$=3.973356 AU

Table 7. The results of approximation of evolution of semimajor axis of fictive orbits in vicinity of the 3:2 resonance. Planet eccentricity is equal eJ=0.048, planet mass is equal m=0.01J

| File | Semimajor axis approximation | sigma |
|---|---|---|
| 236 | A=4.005+0.00122*cos(20.921t+ 1.01) | 0.000407 |
| 235 | 3.9943+0.001664*cos(14.371t+0.996) | 0.000441 |
| 234 | 3.9887+0.002258*cos(11.20t+0.996) | 0.000649 |
| | | |
| 226 | 3.9525+ 0.00200*cos(11.260t+2.50) | 0.000499 |
| 225 | 3.9420+0.001313*cos(17.718t+2.99) | 0.000493 |
| 224 | 3.9368+0.00110*cos(20.70t+2.00) | 0.000422 |

The approximating equations are:
F= -603.27 $a$ + 2395.7
F= 598.47$a$ - 2376
The corresponding intersection point ordinate is $a$=3.97066 AU

We belief that the position of intersection point is determined the position of the resonance. As a result, we note, that the position of resonance moved from nominal in a direction of larger semimajor axis when the perturbing planet mass increase. In the same time, the value of frequency at the intersection point increases with the planet mass.

Table 8. The results of approximation of evolution of semimajor axis of fictive orbits in vicinity of the 3:2 resonance. Planet eccentricity is equal eJ=0.048, planet mass is equal m=0.5J

| File | Semimajor axis approximation | sigma |
|---|---|---|
| 238 | 4.101+ 0.00864cos(0.0747t+0.00) | 0.0097 |
| 242 | 4.065+0.01440cos(0.054246t+0.101) | 0.0073 |
| 237 | 4.0425+0.0190cos(0.04230t+0.996) | 0.0067 |
| | | |
| 225 | 3.945+0.0150cos(0.01882t+1.400) | 0.0043 |
| 224 | 3.9405+0.01344cos(0.020741t+1.998) | 0.0032 |
| 223 | 3.937+ 0.01164cos(0.022655t+ 1.005) | 0.0049 |

The approximating equations are:
F= -476.93x + 1900.2
F= 555.3x - 2202.7
The corresponding intersection point ordinate is $a$=3.974792 AU

**3:1 resonance**

The value of semimajor axis corresponded with exact 3:1 commensurability with Jupiter is $a_{res}$=2.50114 AU. As it is evident in the fig 11, the frequency of the resonant perturbations decreases in approach to resonance when the amplitude increases.

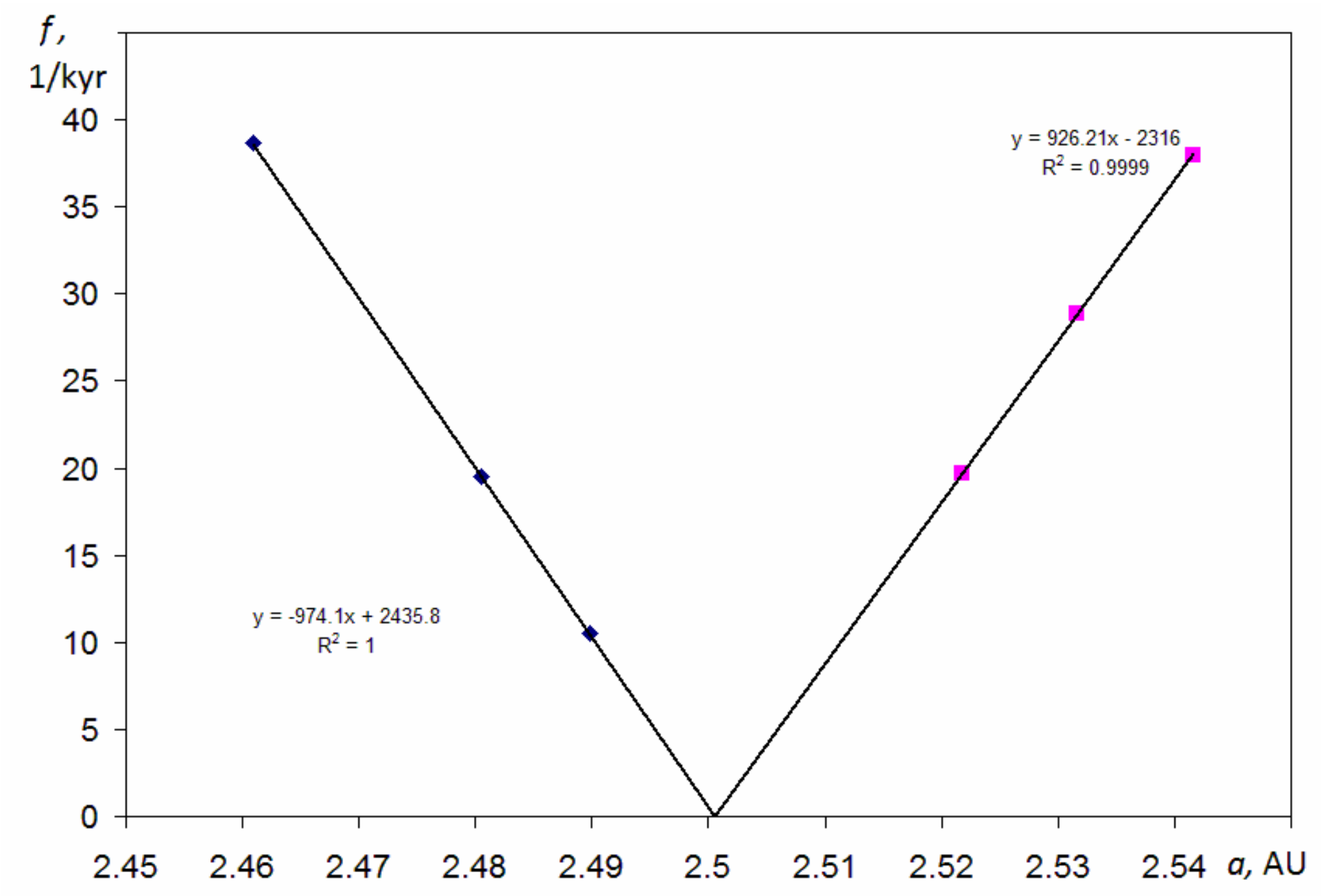

Fig.11. The dependence of the resonant perturbation frequency on the semimajor axis near to 3:1J resonance. The nominal position of resonance is equal 2.50114 AU.

Below we calculate the coordinates of the intersection points of approximating lines and the ratio of the coefficients of approximated lines above and below the resonance( ratio-/+).

Table 9. The results of approximation of evolution of semimajor axis of fictive orbits in vicinity of the 3:1 resonance. Planet eccentricity is equal eJ=0,

a) planet mass is equal m=0.1J

| File | Semimajor axis approximation |
|---|---|
| 329 | 2.551282+0.000040cos(0.04636t+5.0401) |
| 328 | 2.54128+0.000053cos(0.037163t+4.7471) |
| 327 | 2.531271+0.000073cos(0.028044t+5.050) |
| 326 | 2.52125+0.000115cos(0.018756t+5.0754) |
| 316 | 2.49144+0.000224cos(0.009409t+4.1041) |
| 315 | 2.48138+0.000106cos(0.019112t+4.0401) |
| 314 | 2.46135+0.000045cos(0.038761t+ 4.008) |
| 313 | 2.47136+0.000066cos(0.02890t+4.0401) |

y = -0.9756x + 2.44=0.9183x - 2.2966

Ratio-/+=1.062398

The corresponding intersection point ordinate is *a*=2.5009768 AU

b) planet mass is equal m=0.5J

| File | Semimajor axis approximation | sigma |
|---|---|---|
| 329 | 2.55105+0.000206cos(0.046476+ 5.478) | 0.0000236 |
| 328 | 2.54098+0.000275cos(0.03720t+ 5.025) | 0.0000264 |
| 327 | 2.53094+0.000369cos(0.027944t+5.025) | 0.0000326 |
| 326 | 2.5208+0.000608cos(0.01860t+5.3106) | 0.0000161 |
| 316 | 2.49185+0.001158cos(0.008779t+ 4.020) | 0.0000334 |
| 315 | 2.48150+0.000525cos(0.018744t+4.020) | 0.0000234 |
| 314 | 2.46137+0.000222cos(0.038627t+4.3824) | 0.0000195 |
| 313 | 2.47140+0.000320cos(0.028779t+4.5816) | 0.0000389 |

y = -0.9806x + 2.4523=0.9216x - 2.3045

ratio-/+=1.064019

The corresponding intersection point ordinate is *a*=2.50068342 AU

c)planet mass is equal m=1J

| File | Semimajor axis approximation | sigma |
|---|---|---|
| 329 | 2.55069+0.000480cos(0.046288t+5.025) | 0.0000548 |
| 328 | 2.54060+0.000504cos(0.037225t+5.478) | 0.0000855 |
| 327 | 2.53050+0.000798cos(0.027755t+5.126) | 0.0000520 |
| 315 | 2.48160+0.001082cos(0.018332t+4.104) | 0.0000507 |
| 314 | 2.46138+0.000480cos(0.038200t+4.012) | 0.0000339 |
| 313 | 2.47145+0.000640cos(0.028253t+ 4.02) | 0.0000351 |

The nominal position of resonance is equal 2.5011433 AU.

y = -0.9826x + 2.4567=0.9179x - 2.295

ratio-/+=1.070487

The corresponding intersection point ordinate is *a*=2.50023678AU.

Note that the ratio of the coefficients of approximated lines is about the same for the different values of perturbing mass and is equal about 1.065±0.005. It means that asymmetry is still present in the case of second order resonance.

**Notes to the elliptic three body problem**

A mean motion resonance between an asteroid and j-th planet occur, when $kn - k_j n_j = 0$ where k and $k_j$ are positive integers, $n_j$ is the mean motion frequency of j-th planet and n is the mean motion frequency of asteroid.

$$k / k_j = n_j / n = 0$$

If $k_j$< k (configuration Sun-Planet-Asteroid) then resonance called exterior (outer), if $k_j$> k (configuration Sun-Asteroid-Planet) then resonance called interior (inner).

Let us to consider the interior resonance. The equation for the critical resonant argument in planar case is:

$$\varphi = j_1 n_p - j_2 n + j_4 \dot{\varpi} = 0 \qquad (3)$$

The secular variations of eccentricity are determined by expression:

$$e = e_0 + \frac{n\alpha}{\dot{\varpi}}(m_p / m_S) C_3 e_p (\cos\dot{\varpi}t - \cos\varpi_0)$$

It is evident, that in case elliptic thee body problem the secular perturbations of *a* and *e* have the equal frequency and may be easy detected (see Tables 1,5,6). However, in case elliptic thee body problem the secular perturbations depending on $\dot{\varpi}$ appear.

Therefore it is necessary to calculate the value of perihelion rotation rate $\dot{\varpi}$ . As it is known for non-resonant case (Murray,Dermott,1999):

$$\dot{\varpi} = n\alpha(m_p / M)\lfloor 2C_1 + C_3 e_p / e\cos(\varpi)\rfloor \qquad (16)$$

where:

$$C_1 = \frac{1}{8}\left[2\alpha\frac{d}{d\alpha} + \alpha^2\frac{d^2}{d\alpha^2}\right] b_{1/2}^{(0)}(\alpha) \qquad (17)$$

$$C_3 = \frac{1}{4}\left[2 - 2\alpha\frac{d}{d\alpha} + \alpha^2\frac{d^2}{d\alpha^2}\right] b_{1/2}^{(1)}(\alpha)$$

Here $b_{1/2}^{(k)}(\alpha)$ – Laplace coefficients, $\alpha = a / a_p$ is the ratio of semimajor axis asteroid and planet. Index p denotes planet orbital elements.

The value of resonant perihelion rate is (Murray, Dermott, 1999, equation (8.30), table 8.1):

$$\dot{\varpi} = 2C_1 + |j_4| C_r e^{|j_4|-2} \overline{\cos\varphi} \qquad (23a)$$

$$C_r = \frac{1}{2}\left(\frac{m_p}{M}\right) n\alpha\left[-2j_1 - \alpha\frac{d}{d\alpha}\right] b_{1/2}^{(j_1)}(\alpha) \qquad (24a)$$

The well known fact that perihelion in resonance has reverse rotation can be used for alternative estimation of the resonance width. At the fixed test particle eccentricity (e=0.1), the region where $\dot{\varpi}$ is negative is narrow (Fig. 12) and it increases with mass of perturbing planet. We outline the notable asymmetry of this region relative nominal position of resonance.

As we note above, the resonance 3:2J with Jupiter can overlap with 4:3J resonance. According to [1], the overlapping can take place since e=0.15. Our integration of fictive asteroid with initial semimajor axis close to 4.1 AU under the perturbation of planet with real Jupiter mass shows a very unstable behavior (Fig.13).

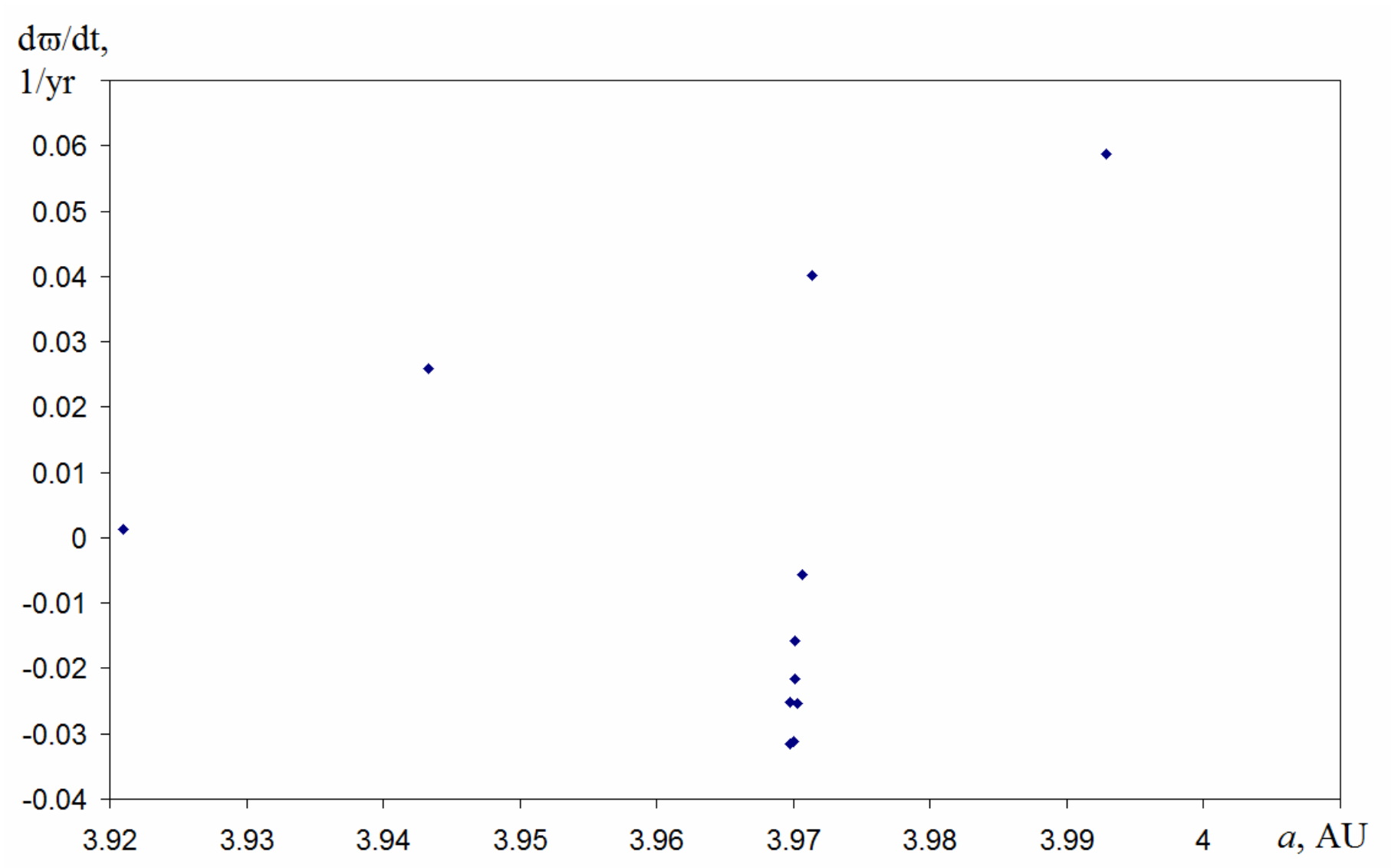

Fig. 12. The perihelion rate on semimajor axis dependence. Planet mass is equal m=0.1 $m_J$.

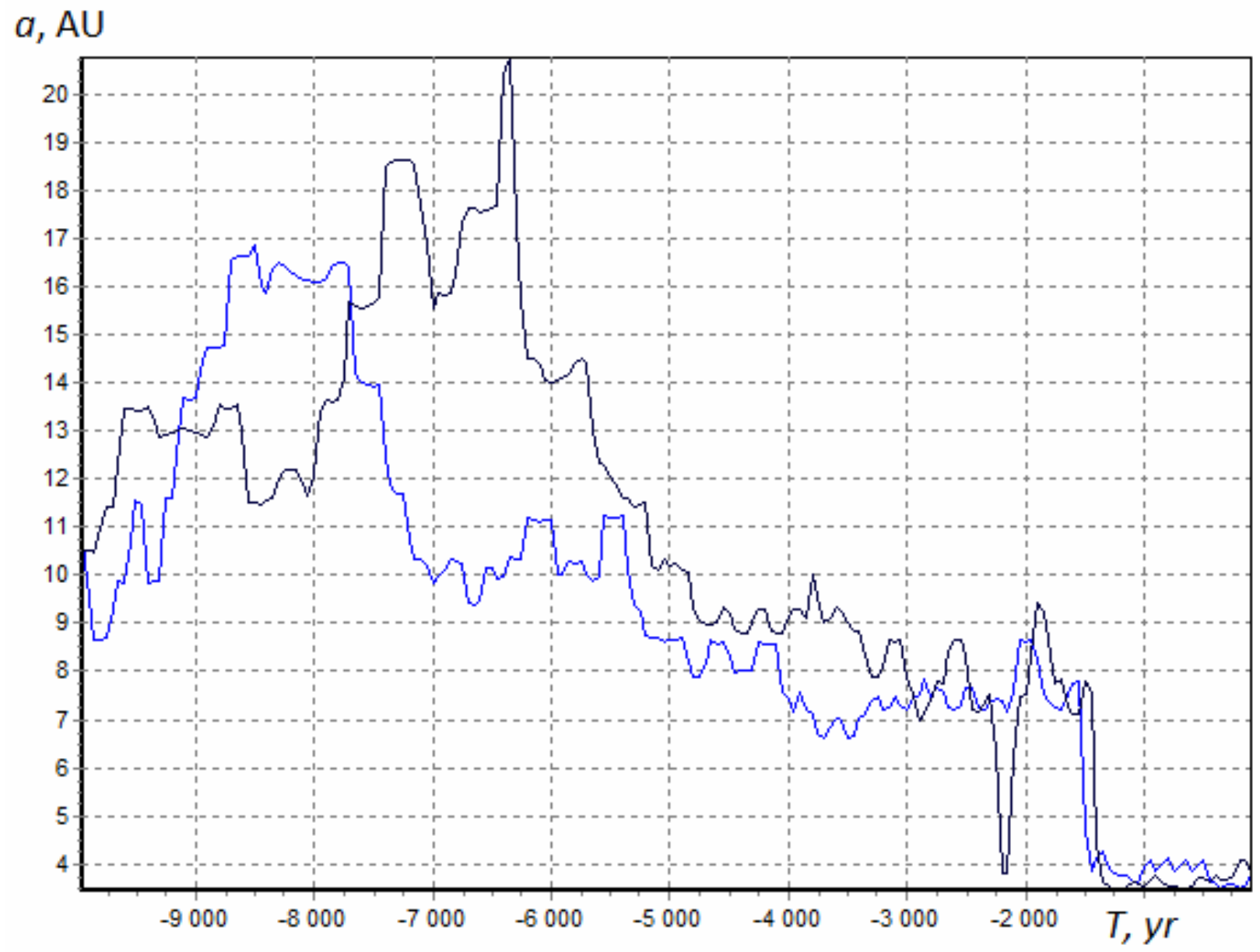

Fig.13. The semimajor axis of test particle evolution between 3:2 and 4;3 resonances with Jupiter

**Conclusions**

As we note above, the main target of this preprint is to give data for future theoretic study. Some of features of resonant motion are noted in our previous papers. In particular, linear dependence of frequency of resonant perturbations on the distance to resonance is reported in [7], [8] for case weak resonances. Here we add consideration by main resonances first and second order in asteroid belt.

The mail conclusions of our studying are:

- The perturbation frequency linearly decreases in approach to exact resonance. The effect can be used to estimate the position of the center of resonance.

- The values of resonance related frequencies are the same both for semimajor axis and eccentricity.
- At large mass of planet the center of resonance offset from its nominal position.
- All results are true in case elliptic orbit of planet. However, in case elliptic three body problem, in eccentricity additionally present the perturbations, corresponded with perihelion rotation.

**Important, that the asymmetry is still present in the case of second order resonance.**

Additionally, we point attention to the motion between 3:2 and 4:3 resonances and the way to resonant width estimation using the perihelion rotation velocity.